\documentclass[twocolumn,showpacs,preprintnumbers,amsmath,amssymb]{revtex4}

\usepackage{ifpdf} \ifpdf
  \usepackage[pdftex,
   bookmarks,
    pdfstartview={Fit}]
  {hyperref}
\fi
\usepackage{verbatim}

\usepackage{graphicx}
\usepackage{dcolumn}
\usepackage{bm}

\begin{document}

\preprint{APS/123-QED}

\title{Clustering of matter in waves and currents}

\author{Marija Vucelja$^1$}
\author{Gregory Falkovich$^1$}
\author{Itzhak Fouxon$^{1,2}$}
\affiliation{$^1$ Physics of Complex Systems, Weizmann Institute
of Science, Rehovot 76100, Israel.\\$^{2}$ Racah Institute of
Physics, Hebrew University of Jerusalem, Jerusalem 91904, Israel.}

\date{\today}

\begin{abstract}
The growth rate of small-scale density inhomogeneities  (the
entropy production rate) is given by the sum of the Lyapunov
exponents in a random flow. We derive an analytic formula for the
rate in a flow of weakly interacting waves and show that in most
cases it is zero up to the fourth order in the wave amplitude. We
then  derive an analytic formula for the rate in a flow  of
potential waves and solenoidal currents. Estimates of the rate and
the fractal dimension of the density distribution show that the
interplay between waves and currents is a realistic mechanism for
providing patchiness of pollutant distribution on the ocean
surface.
\end{abstract}

\maketitle

Random compressible flows
produce very inhomogeneous distribution of density, see
\cite{HH,FGV,BFF,R,FF,BGH,BM,BFS,Bruno} for theory and
\cite{gorlum,SO,Sommerer,NAO,alstrom,CG,DFL} for experiments. Here
we study the growth of density inhomogeneities at small scales,
where the flow can be considered spatially smooth. It can then be
characterized by the Lyapunov exponents whose sum is the
logarithmic rate of change of an infinitesimal volume element,
that is minus the density rate of change $\lambda$. It is called
also the entropy production rate or the clustering rate.   Since
contracting regions contain more statistical weight than expanding
ones, $\lambda$ is generally positive in a random compressible
flow \cite{R,BFF,FGV,FF} (an analog of the second law of
thermodynamics). As a result, density grows on most trajectories.
In the limit of infinite time, density concentrates on a
constantly evolving fractal set characterized by a singular
(Sinai-Ruelle-Bowen) measure \cite{BR,R,S,D}. Our goal here is to
establish what determines the rate $\lambda$ in fluid flows with
waves, particularly, on liquid surfaces. Patchiness in the
distribution of litter on the surface of lakes and pools and of
oil slicks and seaweeds on the sea surface is well-known
empirically while there is no theory that describes it. We think
that in many situations patchiness is a signature of a fractal
measure forming on the surface. Our purpose here is to estimate
 how fast this fractal set is formed and what is its fractal
dimension.

Any velocity field can be separated into solenoidal
(incompressible) and potential components. Surface flows can be
compressible, even for incompressible fluids. For example,
underwater turbulence produces compressible surface currents that
lead to fractal distributions of surface density
\cite{CDGS,CG,DFL,NAO,Sommerer,SO}. However, underwater turbulence
is relatively rare in natural environment (because of stable
stratification) and large-scale currents are usually
incompressible. Compressible component of the surface flows is
then provided by potential waves. Linear waves just oscillate, net
effects are produced by nonlinearity. Every running plane wave
provides for a (Stokes) drift proportional to the square of the
wave amplitude. A set of random waves provides for the Lyapunov
exponents proportional to the fourth power of the wave amplitudes
yet the sum of the exponents $-\lambda$ is found to be zero for
purely potential waves with Gaussian statistics (nonzero rate
appears only in the sixth order in wave amplitudes, i.e. it is so
small as to be practically unobservable in most cases) \cite{BFS}.
Here we use the recently derived general formula for the entropy
production rate \cite{FF1} and show that the account of wave
interaction (which makes the statistics weakly non-Gaussian) does
not bring non-zero $\lambda$ in the fourth order in wave
amplitudes.  We then suggest that in many situations (particularly
on liquid surfaces) the growth of density inhomogeneities is due
to an interplay between potential waves and solenoidal currents.
For such flows, we calculate $\lambda$ and the fractal dimension
of the resulting measure  and consider different limits.

In the velocity field ${\bm v}(t,{\bm x})$, the trajectory ${\bm
X}(t,{\bm x})$ satisfies the equation $\partial _t{\bm X}= {\bm
v}(t,{\bm X})$ with the initial condition ${\bm X}(0,{\bm x})={\bm
x}$. The rate of density change along the trajectory averaged over
${\bm x}$ is given by  \cite{FF1}:
\begin{equation} \lambda=-\lim_{t\to\infty}\langle w(t, {\bm X})\rangle =
\int^{\infty} _0\! dt\langle w(0,{\bm x})w (t,{\bm X}) \rangle\
,\label{form1}\end{equation} with $w\equiv {\bm \nabla} \cdot {\bm
v}$. This is a generalization of the Kawasa\-ki formula \cite{KY}
(obtained in the context of statistical physics) to time-dependent
flows with a steady statistics.

For a general flow,
it is impossible to relate the Lagrangian integral (\ref{form1})
to the velocity spectra or correlation functions  given usually in
the Eulerian frame. However, for low-amplitude waves, fluid
 particles shift little during a period, which allows for an analytical treatment.
 Considering
waves with the dispersion relation $\Omega_{\bm k}$ and packets
with both the wavenumber and the width of order $k$, we estimate
the correlation time of $w$ as $\Omega_{\bm k}^{-1}$ and the
correlation length as $k^{-1}$. The deviation $\bm X(t, \bm x)-\bm
x$ during $t\simeq\Omega_{\bm k}^{-1}$ is $\epsilon=kv/\Omega_{\bm k}\ll
1$ times smaller than $k^{-1}$. We now expand (\ref{form1}) near
$\bm x$ up to $\epsilon^4$:
\begin{align}
&\lambda\! \approx \!\int ^\infty_0\!dt\,\biggl[\langle
w(0)w(t)\rangle +\!\langle w(0 ) \frac{\partial w(t)}{\partial
x^\alpha} \int^{t}_0dt_1v^\alpha(t_1 )\rangle \nonumber\\ &
+\left\langle w(0 )\frac{\partial w(t )}{\partial x^\alpha} \int
^{t}_0dt_1\frac{\partial v^\alpha(t_1 )}{\partial x^\beta} \int
^{t_1}_0dt_2v^\beta(t_2 ) \right\rangle
\label{lambda_init_pos_expansion}\\ & + \frac{1}{2}\biggl\langle
w(0 ) {\partial w (t )\over
\partial x^\beta\partial x^\alpha } \int
^{t}_0dt_1v^\alpha(t_1 )\int ^{t}_0dt_2v^\beta(t_2
)\biggr\rangle\biggr]\ .\nonumber
\end{align} All quantities
here are taken at the same point in space.

We start the consideration of (\ref{lambda_init_pos_expansion})
from the simplest case when the flow is solely due to weakly
nonlinear waves. The normal coordinates of such waves satisfy the
equation $\partial _t a_{\bm k}= -i{{\delta \mathcal{H}}/{\delta
a^*_{\bm k}}}$ while the velocity Fourier component is assumed to
be given by ${\bm v}_{\bm k}={\bm A}_{\bm k}(a_{\bm k}-a^*_{-\bm
k})$. Here $\mathcal{H}$ is the wave Hamiltonian, which can be
expanded in wave amplitudes as follows \cite{ZLF}:
 $\mathcal{H}=\int d{\bm k} \Omega _{\bm k}|a_{{\bm k}}|^2
 +{1\over 2}\int d{\bm k}_{123}\left({\bm V}_{123} a_1 a^{\ast} _2
a^{\ast} _3
            +{\rm c.c.}\right)+\cdots\,.$
We do not write explicitly here other (third and fourth-order)
terms since they will not contribute $\lambda$ up to
$\sim\epsilon^4$. We use throughout the shorthand notations ${\bm
V}_{123}=V_{123}\delta ({\bm k}_1-{\bm k}_2-{\bm k}_3)$ and
$V_{123}=V({\bm k}_1,{\bm k}_2,{\bm k}_3)$, {$\Omega(\pm{\bm
k}_i)=\Omega_{\pm i}$} and ${\bm A}({\bm k}_i)={\bm A}_i$.

One derives the clustering rate up to $\epsilon^4$ using a
standard perturbation theory for weakly interacting waves
\cite{ZLF}. The first term in (\ref{lambda_init_pos_expansion}) is
the time integral (the zero-frequency value) of the second moment.
At the order $\epsilon^2$, the second moment in the frequency
representation is proportional to the delta function: $\langle
a^*({\bm k},\omega)a({\bm k}',\omega')\rangle = (2\pi)^{d+2}n({\bm
k})\delta(\omega-\Omega_{\bm k})\delta({\bm k}-{\bm
k}')\delta(\omega-\omega')$ . A finite width over $\omega$ and a
finite value at $\omega=0$ appear either due to finite linear
attenuation (the case considered in  \cite{FS}) or due to
nonlinearity in the second order of perturbation theory (which
gives $\epsilon^4$ and is considered here). The second term  in
(\ref{lambda_init_pos_expansion}) is the triple moment which
appears in the first order of the perturbation theory and the last
two terms contain the fourth moment which is to be taken at the
zeroth order (i.e. as a product of two second moments).
Straightforward calculations then give  for weakly nonlinear waves
the $\epsilon^4$ contribution:

\begin{align} \nonumber
&\lambda =
{\rm Re}\int\frac{d{\bm k}_2d{\bm k}_3}{(2 \pi)^{2d}}
\delta(\Omega_{2}-\Omega_{3})n({\bm k}_2)n({\bm k}_3)
\biggl\{
\int \frac{d{\bm k}_1}{(2 \pi)^{d}}
\\  \label{inter1}
&\times \left[\left(\frac{{\bm V}^*_{213}}{\Omega_1}
\right.-\frac{{\bm V}_{3-12}}{\Omega_{-1}} \right)
\left((2\pi)^{3d+1}|{\bm A}_1\cdot{\bm
k}_1|^2\frac{V_{213}}{\Omega_1} \right.
\\  \label{inter2}
&\left.-\left.\frac{(2\pi)^{2d}}{\Omega _2} ({\bm A}^*_1\cdot{\bm k}_1)({\bm
A}_2\cdot{\bm k}_2) ({\bm A}^*_3\cdot({\bm k}_2+{\bm k}_1))\right)
\right]+
\\ \label{noninter}
& +\frac{\pi}{\Omega ^2_2} \left|({\bm A}_3\cdot{\bm k}_3)({\bm
A}^*_2\cdot{\bm k}_3) -({\bm A}_3\cdot{\bm k}_2)({\bm A}^*_2\cdot{\bm
k}_2)\right|^2 \biggr\}.
\end{align}
The common factor $\delta (\Omega_2 - \Omega _3) {n}({\bm k}_2)
{n}({\bm k}_3)$ tells that we have here the contribution of two
pairs of waves with the same frequencies. All three terms are
generally nonzero (and positive) when the dispersion law is
non-monotonic or non-isotropic so that $\Omega_2=\Omega_3$ does
not require $k_2=k_3$. In most interesting cases, however,
$\Omega_k$ is a monotonous function of the modulus $k$ so that
$k_2=k_3$. Let us show first that wave interaction does not
contribute $\lambda$ in this case. Indeed, the first two terms,
(\ref{inter1},\ref{inter2}), that came out of the first two terms
of (\ref{lambda_init_pos_expansion}), are proportional to the
difference, ${\bm V}^*_{213}-{\bm V}_{3-12}$, between the
amplitude of decay into a wave with ${\bm k}_1$ and confluence
with a wave with $-{\bm k}_1$.  Interaction coefficients for
$k_2=k_3$ have rotational symmetry and are thus functions of
wavenumbers so that
$V_{213}-V_{3-12}=V_{213}-V_{312}=V_{212}-V_{212}=0$.

The last term (\ref{noninter}) comes from the last two terms of
(\ref{lambda_init_pos_expansion}) and does not contain the
interaction coefficient $V$. This term is due to nonlinear
relation between Eulerian and Lagrangian variables rather than due
to wave interaction. We can compare (\ref{noninter}) with the
growth rate of the squared density for non-interacting waves, see
(12) in \cite{BFS} written there in terms of the energy spectrum,
$E^{\alpha \beta}({\bm k},\omega)=2\pi A^\alpha({\bm k})A^{*\beta}({\bm
k})[n({\bm k})\delta(\omega-\Omega_{\bm k})+n(-{\bm k})\delta(\omega+\Omega_{-{\bm k}})]$.
The comparison shows this part of our logarithmic growth rate being
exactly half the growth rate for the second moment
 as it should be for a short-correlated flow
\cite{FGV}.  Indeed, the process of creation of density
inhomogeneities is effectively short-correlated since the time it
takes ($1/\Omega_k\epsilon^4$ or longer) exceeds the correlation
time of velocity divergence in the Lagrangian frame, $1/\Omega_k$.
For monotonous $\Omega(k)$, (\ref{noninter}) is nonzero only if
the polarization vector ${\bm A}_{\bm k}$ is neither parallel nor
perpendicular to ${\bm k}$ i.e. contains both solenoidal and
potential components. This is not the case for most waves in
continuous media. We thus conclude that for most common situations
(in particular, for sound or surface waves) the entropy production
rate is zero in the order $\epsilon^4$. Note that for surface
waves, the canonical variables are elevation $\eta({\bm r},t)$ and
the potential $\phi({\bm r},z=\eta,t)$ which are related to the
surface velocity by a nonlinear relation
 ${\bm v}=\nabla \phi({\bm r},\eta,t)$. Expanding it in the powers
 of $\eta$, one can show
that this extra nonlinearity does not contribute $\lambda$ in the
order $\epsilon^4$ \cite{FV}. We find it remarkable that the flow
of random potential waves is only weakly compressible (i.e. the
senior Lyapunov exponent is much larger than the sum of the
exponents).

Therefore, we consider now the clustering rate in the presence of
solenoidal currents and potential waves, the situation most
relevant for oceanological applications \cite{P}. Consider the
solenoidal flow $\bm u$  weakly perturbed by potential waves with
$\bm v$. To derive  $\lambda$ in the lowest (second) order in
$\epsilon=kv/\Omega_k$,  we neglect the contribution of $\bm v$
into $\bm X$ in (\ref{form1}) and assume $\partial_t \bm X(t, \bm
x)\approx\bm u(t, \bm X(t, \bm x))$. In this order, $w=\nabla\cdot
\bm v $ is Gaussian and one may integrate by parts: $\langle w(0,
\bm x) w(t, \bm X(t, \bm x))\rangle=\int dt'd\bm x'\Phi(t', \bm
x'-\bm x) \langle \delta w(t, \bm X(t, \bm x)) /\delta w(t', \bm
x')\rangle$. Here $\Phi (t'-t, \bm x'-\bm x)=\langle w(t, \bm x)
w(t', \bm x') \rangle$ is the Eulerian correlation function and
\begin{eqnarray}&&
\lambda\approx\int_0^{\infty} dt\, \langle \Phi[t,  \bm
J(t)]\rangle,\ \ \bm J(t)\equiv \bm X(t, \bm x)-\bm x .
\label{lambda2}\end{eqnarray}  Waves and currents are considered
statistically independent in this order. Using the spectrum,
$k^{\alpha}k^{\beta}E^{\alpha\beta}_{\bm k}\equiv k^2E_{\bm k}$,
we can express $\Phi(t, \bm r)=(2\pi)^{-d}\int k^2E_{\bm
k}\cos(\bm k\cdot \bm r-\Omega_{\bm k} t)d\bm k$ and rewrite
(\ref{lambda2}) as a weighted spectral integral:
\begin{eqnarray}&&\lambda=(2\pi)^{-d}\int
k^2E_{\bm k}\mu(\bm k)\,d{\bm k}\,,\label{lambda23}\\&& \mu(\bm
k)=\int_0^{\infty} \left\langle \cos\left[\bm k\cdot \bm
J(t)-\Omega_{\bm k}t\right]\right\rangle dt.
\label{lambda3}\end{eqnarray} The spectral weight $\mu(k)$
is the Lagrangian correlation time of the
$k$-harmonic of $w$ 
and is expressed via the characteristic function of the particle
drift $\bm J(t)$. Without currents, (\ref{lambda23},\ref{lambda3})
reproduce the first term of (\ref{lambda_init_pos_expansion})
since only the zero-frequency wave contributes.  Already a steady
uniform current $\bar{\bm u}$ contributes the clustering rate in
the order $\epsilon^2$ if there are waves whose Doppler-shifted
frequency is zero in the current reference frame:
$\lambda=(2\pi)^{-d}\int k^2E_{\bm k}\delta (\Omega_{\bm k}-{\bm
k}\cdot \bar{\bm u} )d{\bm k}$. Similar Cherenkov resonance has
been noticed before for diffusivity \cite{Balk}. In the rest of
the paper we consider the fluctuating part of current velocity
characterized  by the rms velocity $u_0^2\equiv \langle
u^2\rangle$ and the correlation time 
$\tau\equiv \int\langle u_{\alpha}(0, \bm x)u_{\alpha}(t, \bm X(t,
\bm x)\rangle dt/u_0^2$. Accordingly, there are two dimensionless
parameters that describe spatial and temporal relationships
between wave and current parameters respectively: $L\equiv
ku_0\tau$ is the ratio between the distance passed by the fluid
particle during $\tau$ and the wavelength, and $T\equiv
\Omega_{\bm k}\tau$ is the ratio between the correlation time of
currents and wave period. The characteristic function $\langle
\exp\left[i\bm k\cdot \bm J(t)\right]\rangle$ generally depends on
the details of the currents statistics but it has universal
behavior both at $t\ll \tau$ and $t\gg \tau$ where general
calculations are possible. On the plane of the dimensionless
parameters $L,T$ we distinguish three regions of different
asymptotic behavior.

Consider first the ballistic limit when the integral
(\ref{lambda3}) is determined by the times $t\ll\tau$ when the
drift velocity does not change and $\bm J(t)\approx \bm u(0, \bm
x)t$. Again, only those waves contribute that are in a Cherenkov
resonance with the current (whose phase velocity coincides with
the local projection of the current velocity):
$\mu=\pi\bigl\langle \delta\bigl(\Omega_{\bm k}-{\bm k}\cdot {\bm
u}\bigr)\bigr\rangle$. In this limit, the weight $\mu$ is
determined by the single-time probability distribution of the
current velocity which we denote ${\cal P}({\bm u})$. In
particular, for the isotropic Gaussian ${\cal P}(u)\propto
\exp(-u^2/2u_0^2)$,  we get
\begin{equation}
{\mu}(k)=(\pi
d/2)^{1/2}(ku_0)^{-1}\exp[-(\sqrt{d}\Omega_k/\sqrt2ku_0)^2]\,,
\label{lambda8}\end{equation}  The ballistic approximation and
(\ref{lambda8}) hold when $k^2u_0^2\tau^2/d$ is much larger than
both unity and $\Omega_k\tau$.

The second universal limit is that of a slow clustering which
proceeds for the time exceeding the correlation time of currents.
At $t\gg\tau$, we use the diffusion approximation, $\langle
\exp\left[i\bm k\cdot \bm J(t)\right]\rangle=\exp[-k^2u_0^2\tau
t/d]$, in (\ref{lambda3}):
\begin{eqnarray}&&
{\mu}(k)=\tau\frac{d(ku_0\tau)^2}{(ku_0\tau)^4+(d\Omega_k\tau)^2}\
. \label{lambda9}\end{eqnarray} That answer and the diffusive
approximation hold when both $k^2u_0^2\tau^2/d$ and $\Omega_k\tau$
are small. Formulas (\ref{lambda23},\ref{lambda9}) can be compared
with the expression for the clustering rate for waves with a
linear damping, $\lambda\simeq \int
k^2E_k\gamma_k(\Omega_k^2+\gamma_k^2)^{-1}d{\bm k}$ \citep{FS}. We
see that in this limit the diffusive motion of fluid particles due
to currents is equivalent in its effect to a damping of waves with
$\gamma_k=k^2u_0^2\tau/d$, where $u_0^2\tau /d$ is the eddy
diffusivity.

The third asymptotic regime takes place for fast-oscillating waves
when $\Omega_k\tau$ exceeds both unity and $k^2u_0^2\tau^2/d$. An
integral of the fast oscillating exponent with a slow function,
$\int_0^\infty\cos(\Omega_kt)f(t)\,dt$, decays as $\Omega_k^{-2n-2}$
where $2n+1$ is the lowest order of the non-vanishing derivative of
$f(t)=\langle \exp\left[i\bm k\cdot\bm J(t)\right]\rangle$  at
$t=0$. When all odd derivatives at zero are zero, the integral
decays exponentially. We see that the answer depends on the details
of the statistics of currents.

If ${\bm u}(t,{\bm X}(t,{\bm x}))$ is Gaussian and isotropic with
$\langle u^\alpha (0,{\bm x})u^\beta (t,{\bm X}(t,{\bm x}))\rangle
=(u^2_0/d)\delta ^{\alpha\beta}\exp(-|t|/\tau)$ then
\begin{equation}
\mu(k)\!=\tau\!\int_0^\infty\!\!\!\! ds \cos(Ts)
\exp\left[(L^2/d)\left(1-s -e^{-s}\right)\right]. \label{fig1}
\end{equation}
It gives both limits (\ref{lambda8},\ref{lambda9}) and
\begin{equation}
\mu(k)=(ku_0)^2/\tau \Omega_k^{4}d\,,\label{fastexp}\end{equation}
at large $\Omega_k$ since the lowest non-vanishing derivative is
$f'''(0)$. Isolines of (\ref{fig1}) are shown in Figure 1 for
arbitrary parameters. Remind that the whole description based on
(\ref{lambda2}) is valid when $v\ll u$.

\begin{figure}
  \includegraphics[width=3in,angle=0]{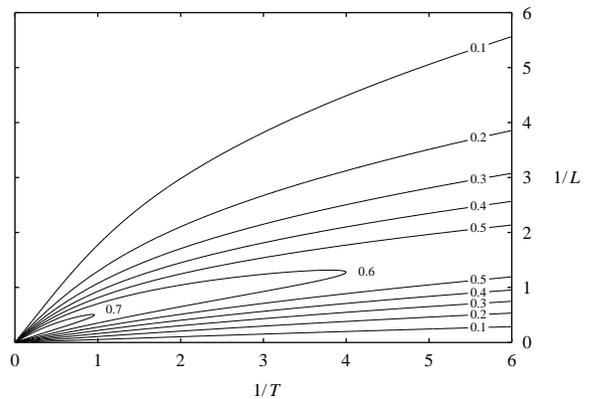}
  \caption{\label{fig:lambda_gaussian}The
isolines of the dimensionless clustering rate $\mu (k) \Omega_{\bm
k}$ given by (\ref{fig1}), here
$ {L}=ku_0\tau$, $ {T}=\Omega_{\bm k}\tau$.}
\end{figure}

If one interpolates between the ballistic  and diffusive regimes
(i.e. between $J^2\propto t^2$ and  $J^2\propto t$) with the help
of the function $\sqrt{1+(t/\tau)^2}-1$, which is smooth at $t=0$,
then the weight factor can be calculated analytically:
\begin{align} \nonumber
&\mu(k)\!=\tau\!\int_0^\infty\!\!\!\! ds \cos(Ts)
\exp\left[(L^2/d)\left(\sqrt{1+s^2}-1\right)\right].
\\\label{K}
& =\frac{\tau L^2}{d}\exp\left(\frac{L^2}{d}\right)
\frac{K_1\left(\sqrt{(L^4/d^2)+T^2}\right)}{\sqrt{L^4/d^2+T^2}}\,,
\end{align}
Here $K_1(x)$ is a Bessel function of an imaginary argument having
the following asymptotics:
$K_1(x)=\sqrt{\pi/2x}\exp(-x)[1+\mathcal{O}(1/x)]$ for $x\gg1$ and
$K_1(x)\simeq 1/x+\mathcal{O}(x\ln(x))$ for $x\ll1$. We see that
(\ref{K}) reproduces  (\ref{lambda8}, {\ref{lambda9}) in the regions
$L^2/d\gg 1$, $L^2/d\gg T$ and $L^2/d\ll 1$, $T\ll 1$ respectively.
At the fast-oscillation limit one gets an exponentially small
contribution
\begin{eqnarray}&&
{\mu}(k)=\tau\sqrt{\frac{\pi
(ku_0)^4\tau}{2d^2\Omega_k^3}}e^{-\Omega_k\tau}\ .\label{fast}
\end{eqnarray}
That concludes the analysis of the weight $\mu(k)$ and we can now
turn to (\ref{lambda23}) to get the clustering rate $\lambda$.

When  the wave spectrum is not very wide (with the width
comparable to $k$) we get
\begin{equation}\lambda\simeq (kv)^2\mu(k)=\epsilon^2\Omega_k^2\mu(k)\
. \label{lambda0}\end{equation} Let us now find out which
wavenumbers contribute (\ref{lambda23}) when the spectrum is wide.
Consider an isotropic power spectrum
$E_{\bm k}\propto 
k^{b-d}$ between some $k_{min}$ and $k_{max}$ and the dispersion
relations $\Omega_{\bm k}=Ck^a$ \cite{ZLF}. Consider first the
ballistic regime. For $(a>1\wedge b>0)\vee(a<1\wedge b<0)$ the
wavenumber $k_*=[bu^2_0/dC^2(a-1)]^{1/(2a-2)}$ determines
$\lambda$. For $(b\leq0\wedge a>1)\vee(b<-1\wedge a=1)$ the
clustering rate is determined by $k_{min}$, while for
$(b\geq0\wedge a<1)\vee (b\geq-1\wedge a=1)$  by $k_{max}$.  Let
us give physical examples using Kolmogorov spectra of waves. For
capillary waves on a deep water, $\Omega_k\propto k^{3/2}$ and
$E_k\propto k^{-11/4}$, and $\lambda$ is determined by $k_{min}$
i.e. by longest waves in the wave turbulent spectrum (assuming the
ballistic approximation is valid for them). For gravity waves on a
deep water, $\Omega_k\propto k^{1/2}$, and for both Kolmogorov
solutions, $E_k\propto k^{-20/6}$ and $E_k\propto k^{-7/2}$, the
clustering rate is determined by waves around $k_*$. For diffusive
regime, the clustering rate is determined by $k_{max}$ if $b\geq
\max[2a-4, 0]$ and by $k_{min}$ if $b< \max[2a-4, 0]$.

Estimates (\ref{lambda8},\ref{lambda9},\ref{fastexp},\ref{fast})
show that $\lambda/(\epsilon^2\Omega_k)\simeq \Omega_k\mu(k)$ is a
dimensionless function which has a maximum of order unity either
in the ballistic regime where the phase velocity of wave is
comparable to the current velocity or in the diffusive regime
where the eddy diffusivity $u_0^2\tau$ is comparable to
$\Omega_{\bm k}k^{-2}$ (in the third asymptotic regime
$\lambda/(\epsilon^2\Omega_k)$ is always small). In those cases,
$\lambda/\Omega_k\simeq \epsilon^2$, i.e. the degree of clustering
during a period is the squared wave nonlinearity (typically
$\epsilon$ is between $0.1$ and $0.01$). Such clustering is pretty
fast (minutes for meter-sized gravity waves and a week for
fifty-kilometer-sized inertio-gravity waves). Therefore, it is
likely that the interplay between waves and currents is the source
of inhomogeneities of floater distribution in many environmental
situations.

Clustering leads to fractal distribution of floaters over the
surface. When compressible component of the velocity is small, the
Lyapunov exponents are due to the current flow,
$\lambda_1\sim\lambda_2\sim\tau^{-1}$. Then, the fractal dimension
of the density distribution can be expressed by the Kaplan-Yorke
formula $1+\lambda_1/|\lambda_2|=2-\lambda/|\lambda_2|\approx
2-\lambda\tau$. The fractal part reaches maximum in the ballistic
regime when $\Omega_k\simeq ku_0$, then $\lambda\tau\simeq
\epsilon^2\Omega_k\tau=\epsilon^2k\ell$ grows with $\ell=u_0\tau$
 and reaches order unity when
$k\ell\simeq \epsilon^{-2}$. Therefore, the distribution is most
fractal when  waves are weak and short while currents are long and
strong. For example, meter-sized gravity waves on water surface
will produce most inhomogeneous distribution of floaters when
there are currents with velocities in meters per second and scales
in hundreds of meters. We see that the current-to-wave ratio of
scales, $k\ell$, compensates for a small wave nonlinearity,
$\epsilon^2$, so that even weak waves with the help of solenoidal
currents can produce very inhomogeneous fractal distribution of
matter.

As a final remark, note that apart from fluid mechanics, one can
think about the evolution of a dynamical system as a flow in the
phase space and treat density as a measure. Solenoidal
(incompressible) flow corresponds to Hamiltonian dynamics and to a
constant (equilibrium) measure. Compressibility corresponds to
pumping and damping i.e. to non-equilibrium. Indeed, the notion of
singular (fractal) measures first appeared in non-equilibrium
statistical physics \cite{BR,S,D} and then was applied in fluid
mechanics \cite{FGV,BFF,BGH,SO,Sommerer}. Therefore, the formulas
(\ref{lambda2}--\ref{fast}) also describe the entropy production
rate in dynamical systems under the action of perturbations
periodic in space and in time.

The work was supported by the ISF. We thank V. Lebedev and E.
Tziperman for helpful explanations.

\bibliography{FFVsub}

\begin{thebibliography}{27}
\expandafter\ifx\csname natexlab\endcsname\relax\def\natexlab#1{#1}\fi
\expandafter\ifx\csname bibnamefont\endcsname\relax
  \def\bibnamefont#1{#1}\fi
\expandafter\ifx\csname bibfnamefont\endcsname\relax
  \def\bibfnamefont#1{#1}\fi
\expandafter\ifx\csname citenamefont\endcsname\relax
  \def\citenamefont#1{#1}\fi
\expandafter\ifx\csname url\endcsname\relax
  \def\url#1{\texttt{#1}}\fi
\expandafter\ifx\csname urlprefix\endcsname\relax\def\urlprefix{URL }\fi
\providecommand{\bibinfo}[2]{#2}
\providecommand{\eprint}[2][]{\url{#2}}

\bibitem[{\citenamefont{Herterich and Hasselmann}(1982)}]{HH}
\bibinfo{author}{\bibfnamefont{K.}~\bibnamefont{Herterich}} \bibnamefont{and}
  \bibinfo{author}{\bibfnamefont{K.}~\bibnamefont{Hasselmann}},
  \bibinfo{journal}{J. Phys. Oceanogr.} \textbf{\bibinfo{volume}{12}},
  \bibinfo{pages}{704} (\bibinfo{year}{1982}).

\bibitem[{\citenamefont{Falkovich et~al.}(2001)\citenamefont{Falkovich,
  Gaw\c{e}dzki, and Vergassola}}]{FGV}
\bibinfo{author}{\bibfnamefont{G.}~\bibnamefont{Falkovich}},
  \bibinfo{author}{\bibfnamefont{K.}~\bibnamefont{Gaw\c{e}dzki}},
  \bibnamefont{and}
  \bibinfo{author}{\bibfnamefont{M.}~\bibnamefont{Vergassola}},
  \bibinfo{journal}{Rev. Mod. Phys.} \textbf{\bibinfo{volume}{73}},
  \bibinfo{pages}{913} (\bibinfo{year}{2001}).

\bibitem[{\citenamefont{Balkovsky et~al.}(2001)\citenamefont{Balkovsky,
  Falkovich, and Fouxon}}]{BFF}
\bibinfo{author}{\bibfnamefont{E.}~\bibnamefont{Balkovsky}},
  \bibinfo{author}{\bibfnamefont{G.}~\bibnamefont{Falkovich}},
  \bibnamefont{and} \bibinfo{author}{\bibfnamefont{A.}~\bibnamefont{Fouxon}},
  \bibinfo{journal}{Phys. Rev. Lett.} \textbf{\bibinfo{volume}{86}},
  \bibinfo{pages}{2790} (\bibinfo{year}{2001}).

\bibitem[{\citenamefont{Ruelle}(1999)}]{R}
\bibinfo{author}{\bibfnamefont{D.}~\bibnamefont{Ruelle}}, \bibinfo{journal}{J.\
  Stat.\ Phys.} \textbf{\bibinfo{volume}{95}}, \bibinfo{pages}{21}
  (\bibinfo{year}{1999}).

\bibitem[{\citenamefont{Falkovich and Fouxon}(2004)}]{FF}
\bibinfo{author}{\bibfnamefont{G.}~\bibnamefont{Falkovich}} \bibnamefont{and}
  \bibinfo{author}{\bibfnamefont{A.}~\bibnamefont{Fouxon}},
  \bibinfo{journal}{New J. Physics} \textbf{\bibinfo{volume}{6}}
  (\bibinfo{year}{2004}).

\bibitem[{\citenamefont{Bec et~al.}(2004)\citenamefont{Bec, Gaw\c{e}dzki, and
  Horvai}}]{BGH}
\bibinfo{author}{\bibfnamefont{J.}~\bibnamefont{Bec}},
  \bibinfo{author}{\bibfnamefont{K.}~\bibnamefont{Gaw\c{e}dzki}},
  \bibnamefont{and} \bibinfo{author}{\bibfnamefont{P.}~\bibnamefont{Horvai}},
  \bibinfo{journal}{Phys. Rev. Lett.} \textbf{\bibinfo{volume}{92}},
  \bibinfo{pages}{224501} (\bibinfo{year}{2004}).

\bibitem[{\citenamefont{Balk and McLaughlin}(1999)}]{BM}
\bibinfo{author}{\bibfnamefont{A.}~\bibnamefont{Balk}} \bibnamefont{and}
  \bibinfo{author}{\bibfnamefont{R.}~\bibnamefont{McLaughlin}},
  \bibinfo{journal}{Phys. Lett.~A} \textbf{\bibinfo{volume}{256}},
  \bibinfo{pages}{299} (\bibinfo{year}{1999}).

\bibitem[{\citenamefont{Balk et~al.}(2004)\citenamefont{Balk, Falkovich, and
  Stepanov}}]{BFS}
\bibinfo{author}{\bibfnamefont{A.~M.} \bibnamefont{Balk}},
  \bibinfo{author}{\bibfnamefont{G.}~\bibnamefont{Falkovich}},
  \bibnamefont{and} \bibinfo{author}{\bibfnamefont{M.~G.}
  \bibnamefont{Stepanov}}, \bibinfo{journal}{Phys. Rev. Lett.}
  \textbf{\bibinfo{volume}{92}}, \bibinfo{pages}{244504}
  (\bibinfo{year}{2004}).

\bibitem[{\citenamefont{Eckhardt and Schumacher}(2001)}]{Bruno}
\bibinfo{author}{\bibfnamefont{B.}~\bibnamefont{Eckhardt}} \bibnamefont{and}
  \bibinfo{author}{\bibfnamefont{J.}~\bibnamefont{Schumacher}},
  \bibinfo{journal}{Phys. Rev. E} \textbf{\bibinfo{volume}{64}},
  \bibinfo{pages}{016314} (\bibinfo{year}{2001}).

\bibitem[{\citenamefont{R.~Ramshankar and Gollub}(1990)}]{gorlum}
\bibinfo{author}{\bibfnamefont{D.~B.} \bibnamefont{R.~Ramshankar}}
  \bibnamefont{and} \bibinfo{author}{\bibfnamefont{J.}~\bibnamefont{Gollub}},
  \bibinfo{journal}{Phys. Fluids~A} \textbf{\bibinfo{volume}{2}},
  \bibinfo{pages}{1955} (\bibinfo{year}{1990}).

\bibitem[{\citenamefont{Sommerer and Ott}(1993)}]{SO}
\bibinfo{author}{\bibfnamefont{J.~C.} \bibnamefont{Sommerer}} \bibnamefont{and}
  \bibinfo{author}{\bibfnamefont{E.}~\bibnamefont{Ott}},
  \bibinfo{journal}{Science} \textbf{\bibinfo{volume}{259}},
  \bibinfo{pages}{335} (\bibinfo{year}{1993}).

\bibitem[{\citenamefont{Sommerer}(1996)}]{Sommerer}
\bibinfo{author}{\bibfnamefont{J.~C.} \bibnamefont{Sommerer}},
  \bibinfo{journal}{Phys. Fluids} \textbf{\bibinfo{volume}{8}},
  \bibinfo{pages}{2441} (\bibinfo{year}{1996}).

\bibitem[{\citenamefont{Nameson et~al.}(1996)\citenamefont{Nameson, Antonsen,
  and Ott}}]{NAO}
\bibinfo{author}{\bibfnamefont{A.}~\bibnamefont{Nameson}},
  \bibinfo{author}{\bibfnamefont{T.}~\bibnamefont{Antonsen}}, \bibnamefont{and}
  \bibinfo{author}{\bibfnamefont{E.}~\bibnamefont{Ott}},
  \bibinfo{journal}{Phys. Fluids} \textbf{\bibinfo{volume}{8}},
  \bibinfo{pages}{2426} (\bibinfo{year}{1996}).

\bibitem[{\citenamefont{{Schroder~et~al}}(1996)}]{alstrom}
\bibinfo{author}{\bibfnamefont{E.}~\bibnamefont{{Schroder~et~al}}},
  \bibinfo{journal}{Phys. Rev. Lett.} \textbf{\bibinfo{volume}{76}},
  \bibinfo{pages}{4717} (\bibinfo{year}{1996}).

\bibitem[{\citenamefont{Cressman and Goldburg}(2003)}]{CG}
\bibinfo{author}{\bibfnamefont{J.~R.} \bibnamefont{Cressman}} \bibnamefont{and}
  \bibinfo{author}{\bibfnamefont{W.}~\bibnamefont{Goldburg}},
  \bibinfo{journal}{J. Stat. Phys.} \textbf{\bibinfo{volume}{113}},
  \bibinfo{pages}{875} (\bibinfo{year}{2003}).

\bibitem[{\citenamefont{Denissenko et~al.}(2006)\citenamefont{Denissenko,
  Falkovich, and Lukaschuk}}]{DFL}
\bibinfo{author}{\bibfnamefont{P.}~\bibnamefont{Denissenko}},
  \bibinfo{author}{\bibfnamefont{G.}~\bibnamefont{Falkovich}},
  \bibnamefont{and}
  \bibinfo{author}{\bibfnamefont{S.}~\bibnamefont{Lukaschuk}},
  \bibinfo{journal}{Phys. Rev. Lett.} \textbf{\bibinfo{volume}{97}}
  (\bibinfo{year}{2006}), \eprint{nlin.CD/0511044}.

\bibitem[{\citenamefont{Bowen and Ruelle}(1975)}]{BR}
\bibinfo{author}{\bibfnamefont{R.}~\bibnamefont{Bowen}} \bibnamefont{and}
  \bibinfo{author}{\bibfnamefont{D.}~\bibnamefont{Ruelle}},
  \bibinfo{journal}{Invent.\ Math.} \textbf{\bibinfo{volume}{29}},
  \bibinfo{pages}{181} (\bibinfo{year}{1975}).

\bibitem[{\citenamefont{Sinai}(1999)}]{S}
\bibinfo{author}{\bibfnamefont{Y.~G.} \bibnamefont{Sinai}},
  \bibinfo{journal}{Russian\ Math.\ Surveys} \textbf{\bibinfo{volume}{27}},
  \bibinfo{pages}{21} (\bibinfo{year}{1999}).

\bibitem[{\citenamefont{Dorfman}(1999)}]{D}
\bibinfo{author}{\bibfnamefont{J.}~\bibnamefont{Dorfman}},
  \emph{\bibinfo{title}{Introduction to Chaos in Nonequilibrium Statistical
  Mechanics}} (\bibinfo{publisher}{Cambridge Univ. Press},
  \bibinfo{year}{1999}).

\bibitem[{\citenamefont{Cressman et~al.}(2004)\citenamefont{Cressman, Davoudi,
  Goldburg, and Schumacher}}]{CDGS}
\bibinfo{author}{\bibfnamefont{J.~R.} \bibnamefont{Cressman}},
  \bibinfo{author}{\bibfnamefont{J.}~\bibnamefont{Davoudi}},
  \bibinfo{author}{\bibfnamefont{W.}~\bibnamefont{Goldburg}}, \bibnamefont{and}
  \bibinfo{author}{\bibfnamefont{J.}~\bibnamefont{Schumacher}},
  \bibinfo{journal}{New J. Physics} \textbf{\bibinfo{volume}{6}},
  \bibinfo{pages}{53} (\bibinfo{year}{2004}).

\bibitem[{\citenamefont{Falkovich and Fouxon}()}]{FF1}
\bibinfo{author}{\bibfnamefont{G.}~\bibnamefont{Falkovich}} \bibnamefont{and}
  \bibinfo{author}{\bibfnamefont{A.}~\bibnamefont{Fouxon}},
  \eprint{nlin.CD/0312033}.

\bibitem[{\citenamefont{Yamada and Kawasaki}(1967)}]{KY}
\bibinfo{author}{\bibfnamefont{T.}~\bibnamefont{Yamada}} \bibnamefont{and}
  \bibinfo{author}{\bibfnamefont{K.}~\bibnamefont{Kawasaki}},
  \bibinfo{journal}{Progr. Theor. Phys.} \textbf{\bibinfo{volume}{38}},
  \bibinfo{pages}{1031} (\bibinfo{year}{1967}).

\bibitem[{\citenamefont{Zakharov et~al.}(1992)\citenamefont{Zakharov, L'vov,
  and Falkovich}}]{ZLF}
\bibinfo{author}{\bibfnamefont{V.~E.} \bibnamefont{Zakharov}},
  \bibinfo{author}{\bibfnamefont{V.}~\bibnamefont{L'vov}}, \bibnamefont{and}
  \bibinfo{author}{\bibfnamefont{G.}~\bibnamefont{Falkovich}},
  \emph{\bibinfo{title}{Kolmogorov spectra of turbulence}}
  (\bibinfo{publisher}{Springer-Verlag}, \bibinfo{address}{Berlin},
  \bibinfo{year}{1992}).

\bibitem[{\citenamefont{Falkovich and Shlomo}(2005)}]{FS}
\bibinfo{author}{\bibfnamefont{G.}~\bibnamefont{Falkovich}} \bibnamefont{and}
  \bibinfo{author}{\bibfnamefont{D.}~\bibnamefont{Shlomo}},
  \bibinfo{journal}{Phys.\ Rev.\ E} \textbf{\bibinfo{volume}{71}},
  \bibinfo{pages}{067304} (\bibinfo{year}{2005}).

\bibitem[{\citenamefont{Fouxon and Vucelja}()}]{FV}
\bibinfo{author}{\bibfnamefont{I.}~\bibnamefont{Fouxon}} \bibnamefont{and}
  \bibinfo{author}{\bibfnamefont{M.}~\bibnamefont{Vucelja}}, \eprint{in
  preparation}.

\bibitem[{\citenamefont{Pedlosky}(1987)}]{P}
\bibinfo{author}{\bibfnamefont{J.}~\bibnamefont{Pedlosky}},
  \emph{\bibinfo{title}{Geophysical fluid dynamics}}
  (\bibinfo{publisher}{Springer}, \bibinfo{address}{New York},
  \bibinfo{year}{1987}).

\bibitem[{\citenamefont{Balk}(2005)}]{Balk}
\bibinfo{author}{\bibfnamefont{A.~M.} \bibnamefont{Balk}},
  \bibinfo{journal}{Private communication}  (\bibinfo{year}{2005}).

\end{thebibliography}

\end{document}